\documentclass[french,english,aps,manuscript,preprintnumbers]{revtex4-1}
\usepackage[T1]{fontenc}
\usepackage[latin9]{inputenc}
\usepackage{amstext}
\usepackage{amssymb}
\usepackage{graphicx}
\usepackage{esint}

\makeatletter

\@ifundefined{textcolor}{}
{%
 \definecolor{BLACK}{gray}{0}
 \definecolor{WHITE}{gray}{1}
 \definecolor{RED}{rgb}{1,0,0}
 \definecolor{GREEN}{rgb}{0,1,0}
 \definecolor{BLUE}{rgb}{0,0,1}
 \definecolor{CYAN}{cmyk}{1,0,0,0}
 \definecolor{MAGENTA}{cmyk}{0,1,0,0}
 \definecolor{YELLOW}{cmyk}{0,0,1,0}
}

\makeatother

\usepackage{babel}
\makeatletter
\addto\extrasfrench{%
   \providecommand{\fg}{\ifdim\lastskip>\z@\unskip\fi~\frqq}%
}

\makeatother
\begin{document}

\title{The calculation of the thermal properties of graphene under a magnetic
field via the two-dimensional Dirac oscillator}

\author{Abdelmalek \surname{Boumali}}

\email{boumali.abdelmalek@gmail.com;boumali.abdelmalek@univ-tebessa.dz}

\affiliation{Laboratoire de Physique Appliquée et Théorique, LPAT\\
 Université de Tébessa, 12000, W. Tébessa, Algeria.}

\date{\today}
\begin{abstract}
In this paper, we show, by using the approach of effective mass, that
the model of a two-dimensional Dirac oscillator can be used to describe
the thermal properties of graphene under an uniform magnetic field.
All thermal quantities of graphene, such as the free energy, the mean
energy, the entropy and the specific heat, have been found by using
an approach based on the zeta function.
\end{abstract}
\maketitle

\section{Introduction}

Graphene is a two-dimensional configuration of carbon atoms organized
in a hexagonal honeycomb structure. The electronic properties of graphene
are exceptionally novel. For instance, the low-energy quasi-particles
in graphene behave as massless chiral Dirac fermions which has led
to the experimental observation of many interesting effects similar
to those predicted in the relativistic regime. Graphene also has immense
potential to be a key ingredient of new devices, such as single molecule
gas sensors, ballistic transistors and spintronic devices.

The Dirac relativistic oscillator is an important potential both for
theory and application. It was for the first time studied by Ito et
al \citep{2}. They considered a Dirac equation in which the momentum
$\vec{p}$ is replaced by $\vec{p}-im\beta\omega\vec{r}$, with $\vec{r}$
being the position vector, $m$ the mass of particle, and $\omega$
the frequency of the oscillator. The interest in the problem was revived
by Moshinsky and Szczepaniak \citep{3}, who gave it the name of Dirac
oscillator (DO) because, in the non-relativistic limit, it becomes
a harmonic oscillator with a very strong spin-orbit coupling term.
Physically, it can be shown that the (DO) interaction is a physical
system, which can be interpreted as the interaction of the anomalous
magnetic moment with a linear electric field \citep{4,5} . The electromagnetic
potential associated with the DO has been found by Benitez et al \citep{6}.
The Dirac oscillator has attracted a lot of interest both because
it provides one of the examples of the Dirac's equation exact solvability
and because of its numerous physical applications (see Ref see Ref
\citep{7} and references therein). Recently, Franco-Villafane et
al \citep{8} exposed the proposal of the first experimental microwave
realization of the one-dimensional DO. The experiment relies on a
relation of the DO to a corresponding tight-binding system. The experimental
results obtained, concerning the spectrum of the one-dimensional DO
with and without the mass term, are in good agreement with those obtained
in the theory. In addition, Quimbay et al \citep{9,10} show that
the Dirac oscillator can describe a naturally occurring physical system.
Specifically, the case of a two-dimensional Dirac oscillator can be
used to describe the dynamics of the charge carriers in graphene,
and hence its electronic properties. Also, the exact mapping of the
DO in the presence of a magnetic field with a quantum optics leads
to regarding the DO as a theory of an open quantum systems coupled
to a thermal bath \citep{11}.

The particle effective mass in graphene is a challenging concept because
the commonly used theoretical expression is mathematically divergent.
Ariel et al \citep{12,13} use a basic principles to present a simple
theoretical expression for the effective mass that is suitable for
both parabolic and non-parabolic isotropic materials such a graphene.
In particular, they demonstrate that the definition of effective mass
is consistent with the definition of the cyclotron effective mass,
which is one of the common methods for effective mass measurement
in solid state materials, and consequently their proposed definition
of the effective mass can be used for non-parabolic materials such
as graphene.

In this article, we attempt to introduce the Dirac oscillator interaction
for describing the electronic properties of graphene in an external
magnetic field, by using the concept of the effective mass. This model
leads to the relativistic dispersion relation observed for graphene,
and explain the existence of a chiral phase transition. Also, it allows
us to investigate the thermodynamic properties of graphene in the
presence of a constant magnetic field. In particular, we determined
the behavior of the main thermodynamical functions: the free energy,
the mean energy, the entropy and the specific heat.

This work is outlined as follows. In Sec. II, we expose our approach
concerning the introduction of the Dirac oscillator interaction by
using the formalism of the effective mass in graphene: in this case,
the eigensolutions have been obtained by using the formalism of a
complex (two-dimensional) quantum relativistic harmonic oscillator.
In Sec. III, the all thermodynamic quantities that describe the thermal
physics of graphene have been calculated via an approach based on
zeta function. Finally, in sec. IV, we present the conclusion.

\section{Eigensolutions of graphene via a two-dimensional massless Dirac oscillator}

\subsection{complex formalism}

In terms of complex coordinates and its complex conjugate \citep{14}
, we have 
\begin{equation}
z=x+iy,\,\bar{z}=x-iy,\label{eq:1}
\end{equation}
and
\begin{equation}
\frac{\partial}{\partial z}=\frac{1}{2}\left(\frac{\partial}{\partial x}-i\frac{\partial}{\partial y}\right),\,\frac{\partial}{\partial\bar{z}}=\frac{1}{2}\left(\frac{\partial}{\partial x}+i\frac{\partial}{\partial y}\right).\label{eq:2}
\end{equation}
The operators momentum $p_{x}$ and $p_{y}$, in the Cartesian coordinates,
are defined by
\begin{equation}
p_{x}=-i\hbar\frac{\partial}{\partial x},\, p_{y}=-i\hbar\frac{\partial}{\partial y}.\label{eq:3}
\end{equation}
When we use $p_{z}=-i\hbar\frac{\partial}{\partial z}$, we get
\begin{equation}
p_{z}=-i\hbar\frac{d}{dz}=\frac{1}{2}\left(p_{x}-ip_{y}\right),\label{eq:4}
\end{equation}
\begin{equation}
\bar{p}_{z}=-i\hbar\frac{d}{d\bar{z}}=\frac{1}{2}\left(p_{x}+ip_{y}\right),\label{eq:5}
\end{equation}
with $p_{z}=-\bar{p}_{z}$. These operators obey the basic commutation
relations
\begin{equation}
\left[z,p_{z}\right]=\left[\bar{z},p_{\bar{z}}\right]=i\hbar,\,\left[z,p_{\bar{z}}\right]=\left[\bar{z},p_{z}\right]=0.\label{eq:6}
\end{equation}
The usual creation and annihilations operators, $a_{x}$ and $a_{y}$
with
\begin{equation}
a_{x}=\sqrt{\frac{m\omega}{2\hbar}}x+i\frac{1}{\sqrt{2m\omega\hbar}}p_{x},\, a_{y}=\sqrt{\frac{m\omega}{2\hbar}}y+i\frac{1}{\sqrt{2m\omega\hbar}}p_{y},\label{eq:7}
\end{equation}
can be reformulated, in the formalism complex, as follows
\begin{equation}
a_{z}=i\left(\frac{1}{\sqrt{m\omega\hbar}}\bar{p}_{z}-\frac{i}{2}\sqrt{\frac{m\omega}{\hbar}}z\right),\label{eq:8}
\end{equation}
\begin{equation}
\bar{a}_{z}=-i\left(\frac{1}{\sqrt{m\omega\hbar}}p_{z}+\frac{i}{2}\sqrt{\frac{m\omega}{\hbar}}\bar{z}\right).\label{eq:9}
\end{equation}
These operators, also, satisfy the habitual commutation relations
\begin{equation}
\left[a,\bar{a}_{z}\right]=1,\,\left[a,a\right]=0,\,\left[\bar{a}_{z},\bar{a}_{z}\right]=0.\label{eq:10}
\end{equation}

\subsection{The framework theoretical of a two-dimensional Dirac oscillator}

We start with the free massive Dirac equation
\begin{equation}
i\hbar\frac{\partial\psi}{\partial t}=H_{D}\psi,\label{eq:11}
\end{equation}
where the wave function $\psi=\left(\begin{array}[t]{cc}
\psi_{1}, & \psi_{2}\end{array}\right)^{T}$ in the graphene case describes the electron states around the Dirac
point $K$ and $K^{'}$, and the Dirac Hamiltonian is given by
\begin{equation}
H_{D}=c\vec{\alpha}\cdot\vec{p}+\beta mc^{2},\label{eq:12}
\end{equation}
where $\vec{\alpha}$ and $\beta$ are the Dirac matrices.

In the case of graphene, one has massless particles that move through
the honeycomb lattice with a velocity $\tilde{c}\sim1.1\times10^{6}\mbox{m}\mbox{s}^{-1}$
\citep{17}, the so-called Fermi velocity. Thus, E. (\ref{eq:12})
reads
\begin{equation}
H_{D}=\tilde{c}\vec{\alpha}\cdot\vec{p}.\label{eq:13}
\end{equation}

Now, the introduction of the Dirac oscillator interaction in graphene
can be made by using the model of the effective mass $m^{*}$ as follows:
the momentum operator $\vec{p}$, in E. (\ref{eq:13}), could be substituted
by $\vec{p}-im^{\text{*}}\omega\vec{r}$, where the additional term
is linear in r, and $m^{*}$ is the effective mass. In this case,
E. (\ref{eq:11}) becomes 
\begin{equation}
\left[\tilde{c}\sigma_{x}\left(p_{x}-im^{*}\omega x\right)+\tilde{c}\sigma_{y}\left(p_{y}-im^{*}\omega y\right)\right]\psi=E\psi.\label{eq:14}
\end{equation}
With the following definitions of Dirac matrices,
\[
\alpha_{x}=\sigma_{x}=\left(\begin{array}{cc}
0 & 1\\
1 & 0
\end{array}\right),\,\alpha_{y}=\sigma_{y}=\left(\begin{array}{cc}
0 & -i\\
i & 0
\end{array}\right),
\]
E. (\ref{eq:14}) can be decoupled in a set of equations as follows
\begin{equation}
E\left|\psi_{1}\right\rangle =\tilde{c}\left[p_{x}+im^{*}\omega x-ip_{y}+m^{*}\omega y\right]\left|\psi_{2}\right\rangle ,\label{eq:15}
\end{equation}
\begin{equation}
E\left|\psi_{2}\right\rangle =\tilde{c}\left[p_{x}-im^{*}\omega x+ip_{y}+m^{*}\omega y\right]\left|\psi_{1}\right\rangle ,\label{eq:16}
\end{equation}
and so, E. (\ref{eq:13}) reads
\begin{equation}
H_{D}=\left(\begin{array}{cc}
0 & \tilde{c}\left[p_{x}+im^{*}\omega x-ip_{y}+m^{*}\omega y\right]\\
\tilde{c}\left[p_{x}+im^{*}\omega x-ip_{y}+m^{*}\omega y\right] & 0
\end{array}\right).\label{eq:17}
\end{equation}
This last form of Hamiltonian of Dirac can be written, in the complex
formalism \citep{11,15,16}, by 
\begin{equation}
H_{D}=\left(\begin{array}{cc}
0 & 2\tilde{c}p_{z}+im^{*}\omega\tilde{c}\bar{z}\\
2\tilde{c}\bar{p}_{z}-im^{*}\omega\tilde{c}z & 0
\end{array}\right).\label{eq:18}
\end{equation}
In this stage, we can see that the form of the Hamiltonian is modified
according to the sign of the $m^{*}$ as follows:
\begin{itemize}
\item if $m^{*}>0$ , the following operators $2\tilde{c}p_{z}+im^{*}\omega\tilde{c}\bar{z}$
and $2\tilde{c}\bar{p}_{z}-im^{*}\omega\tilde{c}z$ are rewritten
with $\left(a_{z},\bar{a}_{z}\right)$ couple, defined by E. (\ref{eq:7}),
as 
\begin{equation}
2\tilde{c}p_{z}+im^{*}\omega\tilde{c}\bar{z}=2i\tilde{c}^{2}\sqrt{m^{*}\omega\hbar}\left(-i\right)\left(\frac{1}{\sqrt{m^{*}\omega\hbar}}p_{z}+\frac{i}{2}\sqrt{\frac{m^{*}\omega}{\hbar}}\bar{z}\right)=g\bar{a}_{z}\label{eq:19}
\end{equation}
\begin{equation}
2\tilde{c}\bar{p}_{z}-im^{*}\omega\tilde{c}z=-2i\tilde{c}^{2}\sqrt{m^{*}\omega\hbar}\left(i\right)\left(\frac{1}{\sqrt{m^{*}\omega\hbar}}\bar{p}_{z}-\frac{i}{2}\sqrt{\frac{m^{*}\omega}{\hbar}}z\right)=g^{*}a_{z}.\label{eq:20}
\end{equation}

\end{itemize}
So, E. (\ref{eq:17})becomes
\begin{equation}
H_{D}=\left(\begin{array}{cc}
0 & g\bar{a}_{z}\\
g^{*}a_{z} & 0
\end{array}\right),\label{eq:21}
\end{equation}
with $g=2im^{*}\tilde{c}^{2}\sqrt{r}$ is the coupling strength between
orbital and spin degrees of freedom, and $r=\frac{\hbar\omega}{m^{*}\tilde{c}^{2}}$
is a parameter which controls the non relativistic limit. According
to E. (\ref{eq:21}), the Dirac Hamiltonian can be written into another
form as

\begin{equation}
H_{D}=g\left(\sigma^{\dagger}\bar{a}_{z}+\sigma^{-}a_{z}\right),\label{eq:21-1}
\end{equation}
and it correspond to the Anti-Jayne\textquoteright s-Cummings (AJC)
model.
\begin{itemize}
\item if, now, the effective mass $m^{*}<0$, the Dirac Hamiltonian becomes
\begin{equation}
H{}_{D}=\left(\begin{array}{cc}
0 & 2\tilde{c}p_{z}-im^{*^{'}}\omega\tilde{c}\bar{z}\\
2\tilde{c}\bar{p}_{z}+im^{*^{'}}\omega\tilde{c}z & 0
\end{array}\right),\label{eq:22}
\end{equation}

\end{itemize}
with $m^{*^{'}}=-m^{*}>0$. By the same way as in the first case,
the operators $2cp_{z}-im\omega'c\bar{z}$ and $2c\bar{p}_{z}+im\omega'cz$
can be rewritten by
\begin{equation}
2\tilde{c}p_{z}-im^{*^{'}}\omega\tilde{c}\bar{z}=2i\tilde{c}\sqrt{m^{*^{'}}\omega\hbar}\left(-i\right)\left(\frac{1}{\sqrt{m^{*^{'}}\omega\hbar}}p_{z}-\frac{i}{2}\sqrt{\frac{m^{*^{'}}\omega}{\hbar}}\bar{z}\right)=ga_{\bar{z}}\label{eq:23}
\end{equation}
 
\begin{equation}
2\tilde{c}\bar{p}_{z}+im^{*^{'}}\omega\tilde{c}z=-2i\tilde{c}\sqrt{m^{*^{'}}\omega\hbar}\left(i\right)\left(\frac{1}{\sqrt{m^{*^{'}}\omega\hbar}}\bar{p}_{z}+\frac{i}{2}\sqrt{\frac{m^{*^{'}}\omega}{\hbar}}z\right)=g^{*}\bar{a}_{\bar{z}},\label{eq:24}
\end{equation}
when we have introducing a new creation and annihilation operators
$\left(a_{\bar{z}},\bar{a}_{\bar{z}}\right)$ couple, defined as follows
\begin{equation}
a_{\bar{z}}=-i\left(\frac{1}{\sqrt{m^{*^{'}}\omega\hbar}}p_{z}-\frac{i}{2}\sqrt{\frac{m^{*^{'}}\omega}{\hbar}}\bar{z}\right),\,\bar{a}_{\bar{z}}=i\left(\frac{1}{\sqrt{m^{*^{'}}\omega\hbar}}\bar{p}_{z}+\frac{i}{2}\sqrt{\frac{m^{*^{'}}\omega}{\hbar}}z\right).\label{eq:25}
\end{equation}
Thus, the new form of a Dirac Hamiltonian is
\begin{equation}
H_{D}=\left(\begin{array}{cc}
0 & g'a_{\bar{z}}\\
g'^{*}\bar{a}_{\bar{z}} & 0
\end{array}\right).\label{eq:26}
\end{equation}
Here $g'=2im^{*^{'}}\tilde{c}^{2}\sqrt{r'}$ is the coupling strength
between orbital and spin degrees of freedom, and $r^{'}=\frac{\hbar\omega}{m^{*^{'}}\tilde{c}^{2}}$
is a parameter which controls the non relativistic limit. As above
case, E. (\ref{eq:26}) can be transforms into 
\begin{equation}
H_{D}=g\left(\sigma^{\dagger}a_{\bar{z}}+\sigma^{-}\bar{a}_{\bar{z}}\right).\label{eq:27}
\end{equation}
Here we are in the case of Jayne\textquoteright s-Cummings (JC) model.
In both cases $\sigma^{\pm}=\frac{1}{2}\left(\sigma^{x}\pm i\sigma^{y}\right)$
are the spin arising and lowering operators.

We would like mentioned hare two remarks: Firstly, the both cases
described above are in well agreement with the study of quantum phase
transition in the Dirac oscillator performed by Bermudez et al \citep{18}.
Secondly, the relativistic Hamiltonian of a two-dimensional Dirac
oscillator can be mapped onto a couple of Anti-Jayne\textquoteright s-Cummings
and Jayne\textquoteright s-Cummings which describe the interaction
between the relativistic spin or and bosons. This mapped between our
problem in question and quantum optics using trapped ions is realized
experimentally. 

Now, Following Esq. (\ref{eq:15}) and (\ref{eq:16}), the wave functions
$\psi_{1}$ and $\psi_{2}$ can be rewritten in the language of the
complex annihilation-creation operators as
\begin{equation}
\left|\psi_{1}\right\rangle =\frac{g}{E}\bar{a}_{z}\left|\psi_{2}\right\rangle ,\label{eq:28}
\end{equation}
\begin{equation}
\left|\psi_{2}\right\rangle =\frac{g^{*}}{E}a_{z}\left|\psi_{1}\right\rangle .\label{eq:29}
\end{equation}
When we write the component $\left|\psi_{1}\right\rangle $ in terms
of the quanta bases, $\left|n\right\rangle =\frac{\left(a^{\dagger}\right)^{n}}{\sqrt{n!}}\left|0\right\rangle $,
these equations can be simultaneously diagonalized, and the energy
spectrum can be described by \citep{11} 
\begin{equation}
E_{n}^{\pm}=\pm\sqrt{4m^{*}\tilde{c}^{2}\hbar\omega n}.\label{eq:30}
\end{equation}
Now, when we put that $\omega=\frac{\omega_{c}}{2}$ with $\omega_{c}=\frac{eB}{m^{*}}$
\citep{17} is a classical cyclotron resonance, E. (\ref{eq:30})
becomes
\begin{equation}
E_{n}^{\pm}=\pm\sqrt{2}\frac{\hbar\tilde{c}}{l_{B}}\sqrt{n},\label{eq:31}
\end{equation}
with $l_{B}=\sqrt{\frac{\hbar}{eB}}$ is the so-called magnetic length\citep{17}.
This definition of the effective mass is consistent with the definition
of the cyclotron mass, which is commonly used for experimental measurements
of the effective mass. Finally, we apply this definition to graphene
and show that it is in agreement with the experimentally observed
linear dependence between the cyclotron mass and momentum\citep{12,13,17}.
The corresponding total wave function for both positive and negative
eigenstates, after normalization, has the following form
\begin{equation}
\left|\pm E_{n}\right\rangle =\left[\begin{array}{c}
\sqrt{\frac{1}{2}}\left|n\right\rangle \\
\mp i\sqrt{\frac{1}{2}}\left|n-1\right\rangle 
\end{array}\right].\label{eq:32}
\end{equation}
Finally, and according to the Esq. (\ref{eq:32}) and (\ref{eq:33}),
our eigensolutions in the commutative case are in well-agreement with
those obtained by \citep{17}.

Finally, and according to the results found above, a two dimensional
Dirac oscillator can be described graphene under a magnetic field.
This means that the last can be mapped onto a couple of anti-Jaynes-Cummings
and Jaynes-Cummings models, well known in quantum optics.

Now, by using E. (\ref{eq:31}), we are able the calculate all thermal
properties of graphene, such as such as the free energy, the mean
energy, the entropy and the specific heat, have been found by using
an approach based on the zeta function.

\section{Thermal properties of graphene}

\subsection{Methods}

In order to obtain all thermodynamic quantities of the relativistic
harmonic oscillator, we concentrate, at first, on the calculation
of the partition function $Z$. The last is defined by \citep{19}
\begin{equation}
Z=1+\sum_{n=0}^{\infty}e^{-\bar{\beta}\sqrt{n}},\label{eq:33}
\end{equation}
where $\bar{\beta}=\frac{1}{\tau}$, and
\begin{equation}
\tau=\frac{l_{B}}{\sqrt{2}\hbar\tilde{c}}\frac{1}{\beta}=\frac{T}{T_{0}},\label{eq:34}
\end{equation}
with $\tau$ denotes the reduce temperature, and 
\begin{equation}
T_{0}=\frac{\sqrt{2}\hbar\tilde{c}}{l_{B}k_{B}},\label{eq:34-1}
\end{equation}
 is the temperature reference value: when we choose $B=18\mbox{T}$,
the value of this temperature is
\begin{equation}
T_{0}\approx3551K.\label{eq:34-2}
\end{equation}
Using the formula (see Ref \citep{20} and references therein)
\begin{equation}
e^{-x}=\frac{1}{2\pi i}\int_{C}dsx^{-s}\Gamma\left(s\right),\label{eq:36}
\end{equation}
the sum in E. (\ref{eq:33}) is transformed into
\begin{equation}
\sum_{n}e^{-\bar{\beta}\sqrt{n}}=\frac{1}{2\pi i}\int_{C}ds\left(\bar{\beta}\right)^{-s}\sum_{n}n^{-\frac{s}{2}}\Gamma\left(s\right)=\frac{1}{2\pi i}\int_{C}ds\left(\bar{\beta}\right)^{-s}\zeta\left(\frac{s}{2}\right)\Gamma\left(s\right),\label{eq:37}
\end{equation}
with $x=\bar{\beta}\sqrt{n}$, and $\Gamma\left(s\right)$ and $\zeta\left(\frac{s}{2}\right)$
are respectively the Euler and zeta function. Applying the residues
theorem, for the two poles $s=0$ and $s=2$, the desired partition
function is written down in terms of the Hurwitz zeta function as
follows:
\begin{equation}
Z\left(\tau\right)=1+\tau^{2}+\zeta\left(0\right).\label{eq:38}
\end{equation}
Now, using that $\zeta\left(0\right)=\frac{1}{2}$ the final partition
function is transformed into 
\begin{equation}
Z\left(\tau\right)=\tau^{2}+\frac{1}{2}.\label{eq:40}
\end{equation}
From this definition, all thermal properties of our system, such as
the free energy, the entropy, total energy and the specific heat,
can be obtained through the numerical partition function $Z\left(\tau\right)$
via the following relations \citep{19}
\begin{equation}
\bar{F}=\frac{l_{B}}{\sqrt{2}\hbar\tilde{c}}F=-\frac{1}{\bar{\beta}}\ln\left(Z\right)=-\tau\ln\left(Z\right),\bar{U}=\frac{l_{B}}{\sqrt{2}\hbar\tilde{c}}U=-\frac{\partial\ln\left(Z\right)}{\partial\bar{\beta}}=\tau^{2}\frac{\partial\ln\left(Z\right)}{\partial\tau},\label{eq:41}
\end{equation}
\begin{equation}
\bar{S}=\frac{S}{k_{B}}=\bar{\beta}^{2}\frac{\partial\bar{F}}{\partial\bar{\beta}}=\ln\left(Z\right)+\tau\frac{\partial\ln\left(Z\right)}{\partial\tau},\,\bar{C}=\frac{C}{k_{B}}=-\bar{\beta}^{2}\frac{\partial\bar{U}}{\partial\bar{\beta}}=2\tau\frac{\partial\ln\left(Z\right)}{\partial\tau}+\tau^{2}\frac{\partial^{2}\ln\left(Z\right)}{\partial\tau^{2}}.\label{eq:42}
\end{equation}

\subsection{Numerical results and discussions}

\begin{figure}
\includegraphics[scale=0.6]{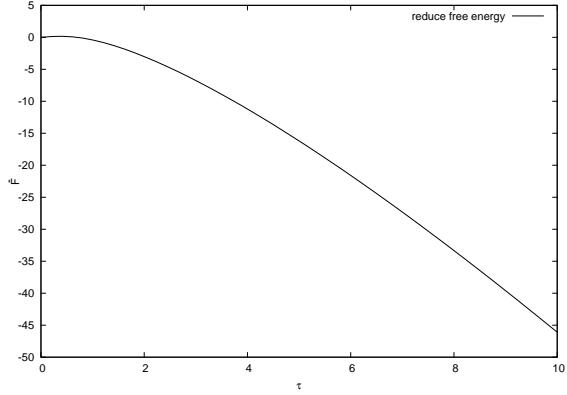}

\protect\caption{\label{fig:1}The reduce free energy $\bar{F}$ versus a reduce temperature
$\tau$.}

\end{figure}

\begin{figure}
\includegraphics[scale=0.6]{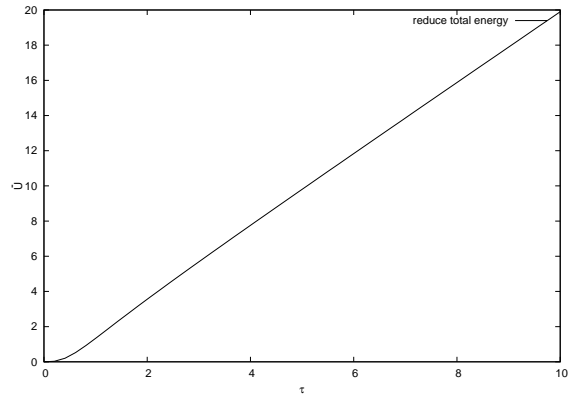}

\protect\caption{\label{fig:2}The reduce total energy $\bar{U}$ versus a reduce temperature
$\tau$.}

\end{figure}

\begin{figure}
\includegraphics[scale=0.6]{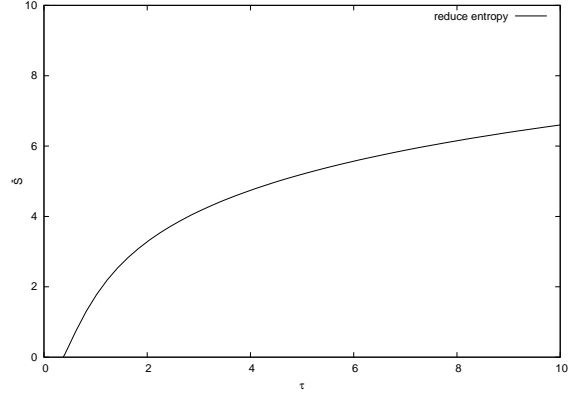}

\protect\caption{\label{fig:3}The reduce entropy function $\bar{S}$ versus a reduce
temperature $\tau$.}

\end{figure}

\begin{figure}
\includegraphics[scale=0.6]{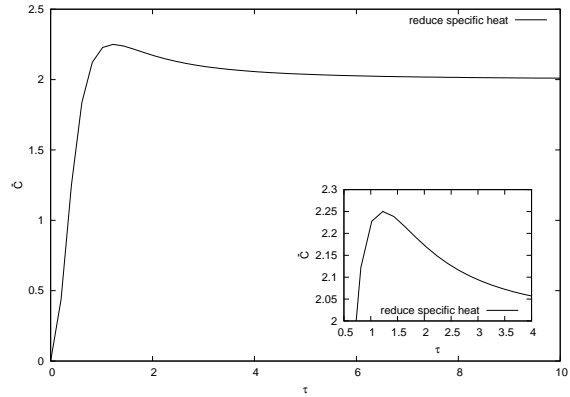}

\protect\caption{\label{fig:4}The reduce specific heat $\bar{C}$ versus a reduce
temperature $\tau$.}

\end{figure}

The thermodynamic quantities are, respectively, plotted in Figures.
\ref{fig:1},\ref{fig:2},\ref{fig:3} and \ref{fig:4}. Before discussing
the main results, we discuss the behavior of the specific heat in
the asymptotic regions, i,e, in the regimes of the higher temperature:
in this region, the partition function can be approximated by
\[
Z\left(\tau\right)\simeq\tau^{2},
\]
which yields to the following results
\[
\bar{U}=2\tau,
\]
\[
\bar{C}=2.
\]
We can argued this by saying that these limits follow the Dulong-Petit
law for an ultra-relativistic ideal gas\citep{19}.

Now, we are ready to present our numerical results for all thermal
properties of graphene under an uniform magnetic field obtained via
our model based on a two-dimensional Dirac oscillator: we should mention
that, in all the figures, we have used adimensional quantities; and
all thermal quantities are plotted versus a reduce temperature $\tau$.

In Fig. \ref{fig:1}, we depicted the free energy versus a reduce
temperature $\tau$: we can see that it decreases with a temperature
as expected. From Fig. \ref{fig:2}, we can see that the total energy
increases with the temperature, and has a linear behavior in higher
temperatures regime. Concerning the entropy function (Fig. \ref{fig:3}),
we observe that it increases with temperature without showing an abrupt
change in its form. That allows us to explain that the curvature,
observed in the specific heat curve (Fig. \ref{fig:4}), does not
exhibit or indicate an existence of a phase transition around a $\tau_{0}=1.5$.
This value of $\tau_{0}$ is different with that found by \citep{19}.

Recently, Sanstos et al \citep{19} have studied the thermodynamics
properties of graphene in non commutative phase-space. They obtained
all thermodynamics quantities by using the Euler-MacLaurin approximation.
We note here, that first, it is not clear why the authors used E.
(10) instead E. (7) in their calculations. Also, the formalism used
by the authors seems to have the following problems: (i) the existence
of the remains term ( Esq. (20) and (21) in Ref \citep{19}) makes
that the calculations become difficult to find, and (ii) the partition
function diverges when $\tau\rightarrow0$. 

Finally, we can see that the thermal properties of graphene can be
obtained easily with the approach based on zeta function than the
formalism of Euler-MacLaurin. Also, we found that our system can be
mapped with quantum optics: the quantum optics theory provides one
of the first testing grounds for the application of the open quantum
systems. The theory of open quantum systems addresses the problems
of damping and dephasing in quantum systems by its assertion that
all real systems of interest are in fact \textquotedblleft open\textquotedblright{}
systems, each surrounded by its environment. Therefore, and following
these arguments, the system of graphene can be regarded as an appropriate
scenario of the theory of an open quantum systems coupled to a thermal
bath (see Ref. \citep{11} and references therein).

\section{Conclusion}

In the present work, we have shown that the formalism of the effective
mass in graphene allowed us to find the eigensolutions of the last
by using the model based on Dirac oscillator interaction. All thermodynamic
properties of graphene have been determined using an approach based
on the zeta function method.


\begin{thebibliography}{99}
\bibitem[1]{1} D. S. L. Abergel , V. Apalkov, J. Berashevich, K.
Ziegler and T. Chakraborty, Advances in Physics, \textbf{59}, 261\textendash 482
(2010).

\bibitem[2]{2}D. Itô, K. Mori and E. Carriere, Nuovo Cimento A, \textbf{51},
1119 (1967).

\bibitem[3]{3}M. Moshinsky and A. Szczepaniak, J. Phys. A: Math.
Gen, \textbf{22}, L817 (1989). 

\bibitem[4]{4}\foreignlanguage{french}{R. P. Martinez-y-Romero and
A. L. Salas-Brito, J. Math. Phys, \textbf{33} , 1831 (1992).}

\bibitem[5]{5}\foreignlanguage{french}{M. Moreno and A. Zentella,
J. Phys. A: Math. Gen, \textbf{22}, L821 (1989). }

\bibitem[6]{6}\foreignlanguage{french}{J. Benitez, P. R. Martinez
y Romero, H. N. Nunez-Yepez and A. L. Salas-Brito,Phys. Rev. Lett,
\textbf{64}, 1643\textendash 5 (1990).}

\bibitem[7]{7}\foreignlanguage{french}{C. Quesne and V. M. Tkachuk,
J. Phys. A: Math. Gen, \textbf{41}, 1747\textendash 65 (2005). }

\bibitem[8]{8}\foreignlanguage{french}{J. A. Franco-Villafane, E.
Sadurni, S. Barkhofen, U. Kuhl, F. Mortessagne, and T. H. Selig- man,
Phys. Rev. Lett. \textbf{111}, 170405 (2013).}

\bibitem[9]{9}{\small{}C. Quimbay and P. Strange, arXiv:1311.2021,
(2013).}{\small \par}

{\small{}\bibitem[10]{10}C. Quimbay and P. Strange, arXiv:1312.5251,
(2013).}{\small \par}

\bibitem[11]{11}\foreignlanguage{french}{{\small{}A. Boumali and
H. Hassanabadi, Eur. Phys. J. Plus. }\textbf{\small{}128}{\small{}:
124 (2013).}}{\small \par}

\bibitem[12]{12}\foreignlanguage{french}{V. Ariel,}{\small{} arXiv:1205.3995,
(2012).}{\small \par}

\bibitem[13]{13}\foreignlanguage{french}{V. Ariel and A. Natan, }{\small{}arXiv:1202.6100v2,
(2012).}{\small \par}

\bibitem[14]{14}\foreignlanguage{french}{A. I. Arbab, Eur. Phys.
Lett, \textbf{98} , 30008 (2012).}

\bibitem[15]{15}\foreignlanguage{french}{B.P. Mandal, S. Verma, Phys.
Lett. A \textbf{374}, 1021 (2010). }

\bibitem[16]{16}\foreignlanguage{french}{B.P. Mandal, S. Verma, Phys.
Lett. A \textbf{376}, 2467 (2012).}

\bibitem[17]{17}\foreignlanguage{french}{Z. Jiang, E. A. Henriksen,
L. C. Tung, Y.-Y. Wang, M. E. Scharwtz, M. Y. Han, P. Kim and H. L.
Stormer, Phys. Rev. Lett. \textbf{98}, 197403, (2007).}

\bibitem[18]{18}\foreignlanguage{french}{A. Bermudez, M. A. Martin-Delgado
and A. Luis, Phys. Rev. A, \textbf{77}, 063815 (2008).}

\bibitem[19]{19}\foreignlanguage{french}{ V. Santos, R.V. Maluf,
C. A. S. Almeida, Annls. of. Phys. \textbf{349},402\textendash 410
(2014).}

\bibitem[20]{20}\foreignlanguage{french}{A. Boumali, arXiv:1409.6205v1
{[}quant-ph{]}, (2014).}

\bibitem[21]{21}E. Elizalde, Ten physical applications of spectral
zeta functions, Springer-Verlag Berlin Heidelberg (1995).\end{thebibliography}
\end{document}